\documentclass[12pt]{article}
\def\bea{\begin{eqnarray}} \def\eea{\end{eqnarray}}

\begin{document}
\title{\bf Eksperimenta havigo de metriko \\ en \^generala relativeco}  
\author{ A.F.F. Teixeira \thanks{teixeira@cbpf.br} \\
         {\small Centro Brasileiro de Pesquisas F\'\i sicas } \\ 
         {\small 22290-180 Rio de Janeiro-RJ, Brazilo} } 
\date{10an de Majo de 2005}
\maketitle 

\abstract 
\^Sajne estas malmulte konata ke la metrikoj de \^generala relativeco (\^GR) povas esti havataj sen integri Einstein-ajn ekvaciojn.  
Por tio, ni bezonas difini nur unuon por \^GR-intertempo $\Delta s$, kaj observi 10 geodezojn (el kiuj, almena\u u unu devas esti nenulan).  
E\^c sen uzi iun unuon, ni povas havi $\kappa g_{\mu\nu}(x^\rho)$, kie $\kappa=$konst.   
Niaj notoj tentas simpligi artikolojn de E. Kretschmann (1917) kaj de H.A. Lorentz (1923) pri tiu lasta afero.   

It seems to be not well known that the metrics of general relativity (GR) can be obtained without integrating Einstein equations. 
To that, we need define only unit for GR-interval $\Delta s$, and observe 10 geodesics (out of which at least one must be nonnull). 
Even without using any unit, we can have $\kappa g_{\mu\nu}(x^\rho)$, where $\kappa=$const. 
Our notes atempt to simplify articles of E. Kretschmann (1917) and of H.A. Lorentz (1923) about this last subject. 
To have an English version of our notes in LaTeX, please ask the author by e-mail.   

\section{Anta\u uparolo} 
Pensu pri finhava 3-spaco (via \^cambro, ekzemple), kaj finhava intertempo (unu minuto komencante nun, ekzemple). 
Tiu duopo estu nomata nia peco de spacotempo.  
Ni scivolas se ni povas findi la \^GR-a metrikon $g_{\mu\nu}(x^\rho)$ en nia peco de spacotempo, sen integri la ekvaciojn de Einstein por la vakua gravita kampo.  

Komence, la \^GR proponas ke la 3-spaco de la \^cambro havu koordinatojn $\{x^i\}$, $i=1..3$;   
tiuj koordinatoj estas ordinare kurbaj.  
Kaj proponas ke, en \^ciu punkto $\{x^i\}$ de la koor\-di\-nat\-sis\-te\-mo, estu fiksata horlo\^go de koordinato, kiu montrus la fluon de la loka tempa koordinato $x^0$. 
Rememoru ke la horlo\^goj de koordinato de la \^GR ne bezonas esti samtempigataj, nek flui same, nek havi konstantan fluon; 
sed \^ciu horlo\^go devas flui \^ciam pozitive, kaj najbaraj horlo\^goj devas \^ciam montri najbarajn tempojn.   

Kiam nia peco de spacotempo jam havas koordinatojn $\{x^\mu\}=\{x^0;\,x^1,x^2,x^3\}$, ni ordinare skribas kaj solvas la ekvaciojn de Einstein, kaj havigas la 10 komponantojn $g_{\mu\nu}(x^\rho)$ de la metrika tensoro.  
Sed ni scivolas se estas ebla, ke ni havigus $g_{\mu\nu}(x^\rho)$ sen solvi tiujn (alte komplikajn) ekvaciojn.  

La respondo estas jes.   
En sekcio 2 ni difinos unuon de tempo kaj de longo; tiuj unuoj estos uzataj por eksperimente havigi la metrikajn komponantojn.   
En sekcioj 3 kaj 4 ni vidos ke, e\^c sen uzi tiuj unuoj, ni eksperimente povas havigi $\kappa g_{\mu\nu}(x^\rho)$, kun $\kappa=$iu nekonata konstanto. 

\section{Sekundo, velo $c$, metro, kaj la $g_{\mu\nu}$'j} 
Nuntempe, {\it sekundo} (unuo de propra intertempo $\tau$) estas difinata \cite{second} kiel la da\u uro de 9.192.631.770 periodoj de la lumo ... inter la du niveloj ... de la atomo de cesio-133.  
La {\it sekundo} estas difinata loke, apud la specimeno de la cesio; 
la difino estas valebla sub multaj kondi\^coj, kiel por iu ajn akcelo de la specimeno \cite{Hay}, kaj por iu ajn pozicio en la gravita kampo.  
 
Anka\u u estas difinata \cite{c} ke la velo de la lumo en la vakuo estas $c$:=299.792.458 m/s. 
Sekve, estas difinata ke {\it metro} estas la longo ke la lumo trakuras en la vakuo dum 1/299.792.458 de sekundo.  
 
Por registri la fluon de la propra tempo (a\u u fizika tempo), atomikaj horlo\^goj estas nuntempe uzataj; 
tiuj horlo\^goj, konsiderataj modelaj, akompanas lineare la fluon de la {\it sekundoj}.   
Ni povas programi modelajn horlo\^gojn por montri la fluon de la produto $c\tau$, anstata\u u de la propra tempo $\tau$. 
Tiu programado estos utila, \^car ni uzos nur la produton $c\Delta\tau $ (anstata\u u $c$ kaj   $\Delta\tau$ malkune) en \^ci tiu artikolo.   

Insistinde, la horlo\^goj de koordinato en \^GR {\it ne} montras lineare la fluon de la fizika tempo, ordinare; ili montras nur la fluon de la loka tempa koordinato $x^0$.  

Posedante modelaj horlo\^goj, ni nun eltrovos la metrikajn koeficientojn $g_{\mu\nu}$ en elekta evento $E$:  
en $E$, ni rimarkas la montrojn $\tau_\alpha \hskip1mm (\alpha=1..10)$ de 10 modelaj horlo\^goj; 
tiuj montroj povas esti malsimilaj, \^ciuj.  
Tuj, ni ek\^\j etas la 10 holo\^gojn la\u u iu ajn direktoj, kun iu ajn veloj.   
Ne longtempe, ni rimarkas la okaza\^{\j}ojn de la 10 horlo\^goj en najbaraj eventoj  $E+\Delta_\alpha  E:=\{x^\mu+\Delta_\alpha x^\mu\}$;  
en tiuj okaza\^{\j}oj ni rimarkas anka\u u la montrojn $\tau_\alpha+\Delta \tau_\alpha$ de la modelaj horlo\^goj. 
Ni skribas 
\bea                                                                \label{10}
c^2\Delta\tau^2_\alpha=g_{\mu\nu}(E)\Delta_\alpha x^\mu\Delta_\alpha x^\nu\,\,, \hskip3mm \alpha=1..10, 
\eea 
kiuj estas 10 ekvacioj, linearaj kaj nehomogenaj, por la 10 nekonatoj $g_{\mu\nu}(E)$. 
Ni solvas tiujn ekvaciojn, kaj havigas la 10 nombrojn $g_{\mu\nu}(E)$. 
Notu en la (\ref{10}) ke, se la modelaj horlo\^goj montrus la fluon de la produto $c\tau$ anstata\u u de $\tau$, tiuokaze la difino de nure {\it metro} estus sufi\^ca por havigi la metrikon.  

Anstata\u u uzi 10 horlo\^gojn, ni povus uzi nur $1\leq n<10$ horlo\^gojn, kaj plus $(10-n$) fotonojn (nulajn geodezojn, $ds^2=0$) kiuj anka\u u pasus tra la evento $E$. 
Tiuokaze la ekvacioj estus    
\bea                                                                \label{20}   
c^2\Delta\tau^2_\alpha&=&g_{\mu\nu}(E)\Delta_\alpha x^\mu\Delta_\alpha x^\nu\,\,, \hskip3mm \alpha=1..n,\\  
0&=&g_{\mu\nu}(E)\Delta_\alpha x^\mu\Delta_\alpha x^\nu\,\,, \hskip3mm \alpha=n+1..10. 
\eea
Tiuj 10 linearaj ekvacioj ($n$ nehomogenaj kaj $10-n$ homogenaj) permesas havigi la 10  $g_{\mu\nu}(E)$. 

Ni bezonus ripeti la procedon por {\it \^ciuj} eventoj de nia peco de spacotempo, por havi la metrikon $g_{\mu\nu}(x^\rho)$ en tiu peco; 
tial ke la nombro de eventoj estas nefinhava, tiu ripetado estas nur formala.  

\section{Havigo de $g_{\mu\nu}(x^\rho)/g_{\sigma\tau}(x^\rho)$} 
E\^c sen uzi unuojn de tempo kaj de longo, ni povas multe scii pri la metrika funkcio  $g_{\mu\nu}(x^\rho)$. 
Frue Kretschmann \cite{Kre} montris kiel havigi $\kappa g_{\mu\nu}(x^\rho)$, kun $\kappa=$ iu ajn konstanto.  
Poste Lorentz \cite{Lor} plibonigis la metodo de Kretschmann.  
\^Ci tie ni resumos la verkon de Lorentz, kun kelkaj modifoj.  

Ni komence elektas eventon $E$, kaj ser\^cas infomojn pri la 10 nombroj $g_{\mu\nu}(E)$. 
En \^stupoj de 1 al 3 ni havigos $\kappa g_{\mu\nu}(E)$, kun $\kappa=$iu ajn konstanto. 

{\bf \^Stupo} 1) Estu fotono kiu trapasas amba\u u $E=\{x^\mu\}$ kaj la najbaran eventon $E_1=E+\Delta_1E$, kun $\Delta_1E=\{\Delta_1x^\mu\}$; estante $ds^2=0$ por fotonoj,   
\bea                                                                   \label{40} 
g_{\mu\nu}(E)\Delta_1x^\mu\Delta_1x^\nu=0.   
\eea 
Tial ke la 4 nombroj $\Delta_1x^\mu$ estas legataj el la koordinatoj, la (\ref{40}) estas 1 homogena lineara ekvacio por la 10 nekonataj nombroj $g_{\mu\nu}(E)$, kun konataj koeficientoj.  

Estu 8 aliaj fotonoj $(i=2...9)$ elektataj hazarde, sed kiuj anka\u u trapasas $E$; 
elektu 8 eventojn $E_2...E_9$, najbarajn al $E$, sur la vojo de la 8 fotonoj; skribu 8 novajn ekvaciojn, linearajn kaj homogenajn, por la 10 nekonatoj $g_{\mu\nu}(E)$. 

{\bf \^Stupo} 2) La aro de la 9 homogenaj linearaj ekvacioj  
\bea                                                                      \label{50}
g_{\mu\nu}(E)\Delta_ix^\mu\Delta_ix^\nu=0, \hskip2mm  i=1..9   
\eea 
donas la valoron de 9 sendependajn raciojn $g_{\mu\nu}(E)/g_{\sigma\tau}(E)$ inter la 10 nombroj  $g_{\mu\nu}(E)$. 
Estus malutile, uzi plus unu fotonon por havi novan ekvacion kiun, kune kun la jamaj 9, permesus havigi la 10 nombrojn $g_{\mu\nu}(E)$; 
tiu nova ekvacio por la $g_{\mu\nu}(E)$'j estus denove homogena, kaj estus dependa de la aliaj ekvacioj.  

{\bf \^Stupo} 3) Ni elektas 10 nombrojn $\kappa^{\sigma\tau}=\kappa^{\tau\sigma}$ kaj difinas  
\bea                                                                       \label{60} 
{\rm e}^{2w(E)}:=\kappa^{\sigma\tau}g_{\sigma\tau}(E), \hskip3mm h_{\mu\nu}(E):={\rm e}^{-2w(E)}g_{\mu\nu}(E):  
\eea 
la nombro $w(E)$ ne estas konata, sed la 10 nombroj $h_{\mu\nu}(E)$ estas \^ciuj konataj.  

{\bf \^Stupo} 4) Ni ripetas la \^stupojn 1..3 por {\it \^ciuj} aliaj eventoj de nia peco de spacotempo, kaj difinas   
\bea                                                                       \label{70} 
{\rm e}^{2w(x^\rho)}:=\kappa^{\sigma\tau}g_{\sigma\tau}(x^\rho), \hskip3mm h_{\mu\nu}(x^\rho):={\rm e}^{-2w(x^\rho)}g_{\mu\nu}(x^\rho);  
\eea
la funkcio $w(x^\rho)$ ne estas konata, sed la 10 funkcioj $h_{\mu\nu}(x^\rho)$ estas \^ciuj konataj. 
Denove, ial ke la nombro de eventoj estas nefinhava, tiu ripetado estas nur formala. 
Rimarku ke la funkcioj $h_{\mu\nu}(x^\rho)$ estis fiksataj per la 10 nombroj $\kappa^{\sigma\tau}$. 
Havante la funkcioj $h_{\mu\nu}(x^\rho)$, ni kalkulas la 40 partajn deriva\^\j ojn $h_{\mu\nu,\lambda}(x^\rho)$, kiuj estos balda\u u uzataj.

Ni ankora\u u bezonas koni la funkcion $w(x^\rho)$ por fine havigi la 10 metrikajn funkciojn $g_{\mu\nu}(x^\rho)$. 
Por tio, Kretschmann kaj Lorentz pensis pri geodezaj movoj de partikloj ($ds^2\neq0$), kiuj ni nun vidos. 
  
\newpage
\section{$w(x^\rho)$ krom obla konstanto} 
Estu $\{x^\mu\}$ kaj $\{x^\mu+\Delta x^\mu\}$ du najbaraj eventoj, amba\u u \^ce geodezo de unu partiklo.    
La movo de la partiklo obeas     
\bea                                                                    \label{80}
\frac{\Delta^2x^\mu}{\Delta s^2}+ \{^{\hskip1mm \mu}_{\nu\rho}\}\frac{\Delta x^\nu}{\Delta s}\frac{\Delta x^\rho}{\Delta s}=0. 
\eea 

Ordinare, la metriko $g_{\mu\nu}(x^\rho)$ uzata en la (\ref{80}) estas konata, kaj la ekvacio donas la movon $x^\mu(s)$ de la partiklo.  
\^Ci tie, tamen, ni uzos la (\ref{80}) aliacele.    
En la 4 ekvacioj (\ref{80}) ni povas noti la $\Delta x^\mu$, sed ni ne konas nek $\Delta s$ (tial ke ni ne uzas unuon de propratempo) nek la simbolojn de Christoffel de dua tipo $\{^{\hskip1mm \mu}_{\nu\rho}\}$ (tial ke ni ne havas nek la funkciojn $g_{\mu\nu}(x^\sigma)$ nek \^giajn deriva\^\j ojn $g_{\mu\nu,\rho}(x^\sigma)$). 
Do la (\ref{80}) estos uzata ne kiel ordinare, sed jes por doni informojn pri la metriko.  

Efektive, la\u ulonge la geodezo ni havas la nombrojn  
\bea                                                                    \label{90}
\Delta\sigma^2:=h_{\mu\nu}(x^\rho)\Delta x^\mu\Delta x^\nu, 
\eea 
tial ke ni havas la $h_{\mu\nu}(x^\rho)$ kaj ni povas noti la $\Delta x^\mu$'s. 
Do ni povas havi, la\u ulonge la geodeza movo, la kvantojn $\Delta x^\mu/\Delta\sigma$ kaj $\Delta^2 x^\mu/\Delta\sigma^2$; tiuj estos uzataj en la (\ref{80}), kun iu modifo. 

Komencu de (\ref{80}) kaj de (\ref{90}), kaj uzu   
\bea                                                                   \label{100}  
\Delta s={\rm e}^w\Delta\sigma, \hskip3mm g^{\mu\nu}=:{\rm e}^{-2w}h^{\mu\nu}, \hskip3mm h^{\mu\nu}h_{\nu\rho}=\delta^\mu_\rho, 
\eea 
\bea                                                                    \label{110}
h_{\mu\nu}\frac{\Delta x^\mu}{\Delta\sigma}\frac{\Delta x^\nu}{\Delta\sigma}=1, \hskip3mm 
\frac{\Delta w}{\Delta\sigma}=\frac{\Delta x^\mu}{\Delta\sigma}w_{,\mu} \hskip1mm,  
\eea 
\bea                                                                    \label{120}
\{^{\hskip1mm \mu}_{\nu\rho}\}_g=\{^{\hskip1mm \mu}_{\nu\rho}\}_h+\delta^\mu_\nu w_{,\rho}+\delta^\mu_\rho w_{,\nu}-h^{\mu\alpha}h_{\nu\rho}w_{,\alpha}\,\, ,  
\eea
por \^san\^gi la (\ref{80}) al   
\bea                                                                   \label{130} 
\left(h^{\mu\alpha}-\frac{\Delta x^\mu}{\Delta\sigma}\frac{\Delta x^\alpha}{\Delta\sigma}\right)w_{,\alpha}=\frac{\Delta^2x^\mu}{\Delta\sigma^2}+\{^{\hskip1mm \mu}_{\nu\rho}\}_h\frac{\Delta x^\nu}{\Delta\sigma}\frac{\Delta x^\rho}{\Delta\sigma}. 
\eea 

En nia peco de spacotempo, elektu geodezo de partiklo, kaj evento $E$ en tiu geodezo. 
\^Ciu kvanto de la (\ref{130}) aludas tiun eventon kaj tiun geodezon.  
Nur la 4 nombroj $w_{,\alpha}(E)$ estas nekonataj, kaj la (\ref{130}) estas 4 algebraj ekvacioj, linearaj kaj nehomogenaj, por tiuj nombroj.  
Tamen, havigi la solvon de la 4 $w_{,\alpha}(E)$ ne estas tiel simpla. Ni vidu kial.    
[\^Ci tie nia metodo malproksimi\^gas al la metodo de Lorentz.] 

Difinu la kvadratan matricon $M$ per \^giaj komponantoj   
\bea                                                                    \label{140}
M^{\mu\alpha}:=h^{\mu\alpha}-u^\mu u^\alpha, \hskip3mm u^\mu:=\frac{\Delta x^\mu}{\Delta\sigma}, 
\eea 
kaj la vertikalan matricon $U$ per  
\bea                                                                    \label{150}
U^\mu:=\frac{\Delta u^\mu}{\Delta \sigma}+\{^{\hskip1mm \mu}_{\nu\rho}\}_h u^\nu u^\rho, 
\eea 
kaj reskribu la (\ref{130}) kiel 
\bea                                                                  \label{160}
M^{\mu\alpha}w_{,\alpha}=U^\mu. 
\eea 
Tial ke la matrico $M$ havas determinanton nulan, 
ni ne povas inversi la (\ref{130}) por havigi $w_{,\alpha}(E)$ simple kiel $(M^{-1})_{\alpha\mu}(E)U^\mu(E)$. 

Tial ni elektas alian geodezon kiun anka\u u trapasus la eventon $E$, sed kun nova velo  $u^{'\mu}(E)$, hazarde elektata.  
Certe la nova ekvacio  
\bea                                                                  \label{170}
M^{'\mu\alpha}w_{,\alpha}=U^{'\mu}
\eea 
anka\u u ne permesas findi $w_{,\alpha}(E)$, tial ke anka\u u la matrico $M^{'}$ havas nulan determinanton.  
Tamen, iu lineara kombino de la ekvacioj (\ref{160}) kaj (\ref{170}), ekzemple la subtraho,   
\bea                                                                 \label{180} 
\left(M^{\mu\alpha}-M^{'\mu\alpha}\right)w_{,\alpha}=U^\mu-U^{'\mu}, 
\eea
permesas havigi la 4 diferencialojn $w_{,\alpha}$ je la evento $E$, 
\bea                                                                 \label{190}
w_{,\alpha}=\left((M-M^{'})^{-1}\right)_{\alpha\mu}(U-U^{'})^\mu,  
\eea
tial ke la determinanto de la matrico $M-M^{'}$ estos nenula.  

Ni bezonas ripeti la kalkulon por \^ciuj eventoj de nia peco de spacotempo, por havi la 4 diferencialajn funkciojn $w_{,\alpha}(x^\mu)$ en la peco.  
Denove tiu ripetado estas nur formala, tial ke la kvanto de eventoj estas nefinhava.   
Pri integrado ni fine havigus la funkcio $w(x^\mu)$, sed kun unu nekonatan adician konstanton  $w_0$. 

{\bf Por fini:} Por havi la valoron de $w_0$, uzu modelan horlo\^gon (prefere unu kiu montru la produton $c\tau$). 
Fiksu tiu horlo\^gon apud iu ajn evento $\{x^\mu\}$ kaj komparu \^gia ritmon al la ritmo de la loka horlo\^go de koordinato (kiu montras $x^0$); skribu  
\bea                                                                \label{200}
c\Delta\tau={\rm e}^{w(x^\mu)}\Delta x^0, 
\eea 
kaj elfindu la valoron de $w_0$. 

\section{Komentoj kaj dankoj} 
Nia metodo por havigi $w_{,\alpha}(E)$, ekvacio (\ref{190}), uzis nur 2 geodezojn ; do \^gi \^sajnas pli simpla ol la metodo de Lorentz \cite{Lor}, kiu uzis 10 geodezojn. 

Konsideru la spacotempo kun koordinatoj $\{x^\mu\}$, metriko $h_{\mu\nu}$, elemento de intertempo $d\sigma$, kaj velo $u^\mu:=dx^\mu/d\sigma$. 
En tiu spacotempo, 
\bea                                                                   \label{210} 
\frac{du^\mu}{d\sigma}+\{^{\hskip1mm \mu}_{\nu\rho}\}_hu^\nu u^\rho=a^\mu, \hskip3mm a^\mu:=\left(h^{\mu\alpha}-u^\mu u^\alpha\right)w_{,\alpha} 
\eea 
estas ekvacio de movo de partiklo, ne geodeza, sed kun 'ekstera akcelo' $a^\mu$. 
Rimarku ke ${M^\mu}_\nu:=\left(h^{\mu\alpha}-u^\mu u^\alpha\right)h_{\alpha\nu}$ faras ${M^\mu}_\nu u^\nu=0$, do \^gi projekcias vektorojn sur la hiperebeno normala al la vektoro de velo $u^\mu$. 
Do la (\ref{130}) permesas havigi nur la komponenton de $w_{,\mu}$ normalan al $u^\mu$, ne la tutan $w_{,\mu}$.  

Ni kore dankas al la skipoj de Microsoft, PCTeX, Reta Vortaro \cite{revo}, kaj arXiv.org, kiuj multe faciligis skribi kaj eldoni \^ci tiujn notojn.

\end{document}